\documentclass[12pt]{article}
\pdfoutput=1
\usepackage{latexsym,cite}
\usepackage{amsmath}
\usepackage{amssymb}
\usepackage{amsthm}
\usepackage{subcaption}
\usepackage{mathtools}
\usepackage{braket}

\usepackage[top = 1. in,bottom = 1 in, left = 1 in, right=1 in]{geometry}

\newcommand{\arXiv}[1]{\href{http://www.arXiv.org/abs/#1}{arXiv:#1}}
\newcommand{\doi}[2]{\href{http://dx.doi.org/#2}{#1}}
\usepackage[colorlinks=true, linkcolor=blue, bookmarks=true]{hyperref}

\allowdisplaybreaks

\makeatletter
\renewcommand\section{\@startsection {section}{1}{\z@}%
                  {-3.5ex \@plus -1ex \@minus -.2ex}
                  {2.3ex \@plus.2ex}%
                  {\normalfont\large\bfseries}}
\renewcommand\subsection{\@startsection{subsection}{2}{\z@}%
                   {-3.25ex\@plus -1ex \@minus -.2ex}%
                   {1.5ex \@plus .2ex}%
                   {\normalfont\bfseries}}
\makeatother

\newcommand{\beq}{\begin{equation}}
\newcommand{\eeq}{\end{equation}}
\newcommand{\del}{\partial}

\newcommand{\ph}{\varphi}
\newcommand{\de}{\delta}
\newcommand{\om}{\omega}
\newcommand{\la}{\lambda}
\newcommand{\ena}{\end{eqnarray}}
\newcommand{\beqa}{\begin{eqnarray}}
\newcommand{\eeqa}{\end{eqnarray}}
\newcommand{\bea}{\begin{eqnarray}}
\newcommand{\eea}{\end{eqnarray}}

\newcommand{\ad}{a^\dagger}
\newcommand{\nn}{\nonumber}

\newcommand{\smallcirc}[1]{\scalebox{#1}{$\circ$}}

\makeatletter
\def\yellowhl#1{{%
  \setlength{\fboxsep}{0pt}%
  \setlength{\fboxrule}{0pt}%
  \color@b@x{\fbox{\phantom{#1}}}%
           {\color[rgb]{1,1,0}\rule[-.2ex]{\wd0}{\ht0}\kern-\wd0 #1}%
}}
\makeatother

\begin{document}
\begin{titlepage}
\begin{flushright}
\phantom{arXiv:yymm.nnnn}
\end{flushright}
\begin{center}
{\huge\bf Spontaneous symmetry breaking\vspace{2mm}\\ on graphs and lattices}
\vskip 10mm
{\large Oleg Evnin}
\vskip 3mm
{\it  High Energy Physics Research Unit, Faculty of Science, \\Chulalongkorn University,
Bangkok, Thailand\vspace{2mm} \\ Theoretische Natuurkunde, Vrije Universiteit Brussel (VUB) \\
\&\,\,
International Solvay Institutes, Brussels, Belgium\vspace{2mm} \\
 Mark Kac Center for Complex Systems Research, Jagiellonian University, Krak\'ow, Poland}
\vskip 7mm
{\small {\tt oleg.evnin@gmail.com}}
\vskip 25mm
{\bf ABSTRACT}\vspace{3mm}
\end{center}
Spontaneous symmetry breaking is a cornerstone of modern physics, defining a wealth of phenomena in condensed-matter and high-energy physics, and beyond.
It requires an infinite number of degrees of freedom, and even then, for continuous symmetries, it only works if the spatial dimension is not too low, following the classic results of Coleman, Hohenberg, Mermin and Wagner. While usually discussed in the context of quantum and statistical field theories, and in particular, effective field theories, there
are advantages in addressing the same kind of phenomena on discrete geometric structures rather than conventional manifolds. When the space is discretized
into a lattice, a lucid picture of conventional spontaneous symmetry breaking springs up, with the ultraviolet issues of continuum quantum field theory out-of-sight, and the key effect, which is infrared in nature, revealed through elementary harmonic oscillator networks. From there, it is natural to generalize lattices to other graphs/networks. In this setting, the presence of spontaneous symmetry breaking is controlled by fractional generalizations of resistance distance and the Kirchhoff index, and most broadly by the spectral dimension.
Predictably, because of the richness of discrete geometric structures in comparison with continuous manifolds, a broader array of geometries emerge where spontaneous breaking of continuous symmetries is blocked by large fluctuations.

\vfill

\end{titlepage}


\tableofcontents\vspace{1cm}

\section{Introduction}

The concept of {\it spontaneous symmetry breaking} (SSB) emerged in the theory of magnetism and superconductivity,
and migrated successfully from there to high-energy physics through the works of Nambu \cite{Nambu}. While what we shall be discussing here
is breaking global symmetries that act uniformly at all spacetime points, akin to spatial rotations being broken by spontaneous magnetization,
the breaking of local gauge symmetries \cite{Kibble,Englert}, akin to the effect of charge condensation in superconductors,
went on to become a pivotal theme in relativistic quantum field theory (QFT) and a key ingredient of the Standard Model of elementary particles \cite{Weinberg}.

The essence of SSB is having a ground state that is not invariant under some of the symmetries of the theory. Evidently, it implies that the ground state is exactly degenerate, since applying a symmetry transformation will produce another equivalent ground state, just like rotating a magnetized sample makes its magnetization point in a different direction. This requires infinitely many degrees of freedom, unless one artificially introduces unphysical infinite energy barriers. Intuitively, with an infinite number of degrees of freedom, some dynamical modes become infinitely massive and get `stuck' at some position in the configuration space, instead of being smeared over all equivalent points connected by symmetries.

Even with an infinite number of degrees of freedom, SSB does {\it not} always happen. The classic results of Mermin-Wagner \cite{MW}, Hohenberg \cite{Hohenberg} and Coleman \cite{Coleman} indicate that when the number of spatial dimensions is small, the long-wavelength fluctuations overpower any possible localization in the field space for the case of SSB of continuous symmetries (see \cite{MaRajaraman} for early historical commentary). Understanding when such phenomena happen on a more broad arena where spatial structures have been replaced by discrete geometries will be the key theme of this essay. 

Similar effects where continuous symmetries cannot be spontaneously broken also emerge in de Sitter spacetimes \cite{Ratra}, in this case, in any number of spatial dimensions. This has led to the concept of quantum de Sitter instability \cite{dSinst,dScosmo} discussed in conjunction with the cosmological constant problem. As to discrete symmetries, while they may be spontaneously broken even in low-dimensional cases, this does not always happen,
and some interesting---and not fully explored---phenomenology emerges \cite{Lukyanov,RychkovVitale}.

Recent years have seen applications of similar ideas to nonrelativistic QFTs \cite{Brauner, WB, Kapustin, WM}, starting with the works by Brauner and Watanabe. The reduced spacetime symmetry creates more options for possible types of SSB. This framework has been later extended to higher-derivative QFTs \cite{GGHY}, yielding an even vaster
array of possible behaviors, and the corresponding extension of the no-SSB results of Coleman-Hohenberg-Mermin-Wagner.

The above classic considerations of SSB are for continuous spacetimes. We will generalize them to discrete spatial structures.
QFTs on discretized spaces have been considered in a number of contexts. The most prominent case is
likely the lattice formulation of gauge theories \cite{Creutz} -- originally for ordinary vector gauge fields, but later also 
for higher-tensor fields \cite{Nepomechie} and topological field theories \cite{CS}. The usual case is on regular lattices,
but random lattices have also been considered \cite{randlat}, paving the way to more general discrete geometries.
Explorations of simplicial quantum gravity \cite{Loll} have provided motivations for studies of quantum fields on simplicial complexes
\cite{smpl,cdt1,cdt2}, and  on more general networks \cite{Filk}, while recent years have seen considerable interest in Dirac operators 
within similar setups \cite{qft1,qft2,qft3,qft4,qft5}, all the way to interacting fermionic QFTs used as probes of the network geometry \cite{netmass}.

In application to SSB, discretization confers a few distinctive advantages, and the purpose of this essay is to explore these advantages.
For conventional SSB in continuous spaces, replacing the geometric background with a hypercubic lattice alleviates all ultraviolet issues typical of QFTs, so that there are no
divergences and the computations are straightforward and well-defined. This is a common rationale for considering latticized QFTs in general.
In our case, it will reduce the question of whether a symmetry is spontaneously broken to an undergraduate-level exercise: computing explicit expectation values in the ground state of a harmonic oscillator network. Pedagogically, this is a very nice way to look at all the standard textbook results in the continuum, as well as their nonrelativistic and higher-derivative generalizations that emerged in recent years \cite{Brauner, WB, Kapustin, WM,GGHY}. From this point, generalization
to arbitrary networks comes in very naturally, since it amounts to replacing lattice Laplacians with graph Laplacians defined in a way that is standard in graph theory and network science. The structure one obtains as a result is richer, however, than what one sees on hypercubic lattices or in continuous spaces, since there is much more room for defining diverse geometries in terms of network connectivities.

Properties of simple QFTs, including SSB, on discrete geometric structures, including networks, are a natural question from a methodological perspective, and it must be settled with clarity.
In applications, such QFTs naturally emerge as a representation for matter fields in discrete approaches to quantum gravity. This most commonly involves simplicial complexes \cite{Loll}, but more general networks have also been considered \cite{combi,birth,combirev}. Speaking of applications to more practical networks science,
in the emerging age of quantum information and quantum communications, multiagent quantum information exchange will lead to communication networks with quantum degrees of
freedom on their nodes, and some of such protocols will be naturally described by QFTs. SSB will then serve as a diagnostic of whether a large network can sustain coherent global order (a collective choice among symmetry-related states) or whether quantum fluctuations wash out such order.

In the exposition that follows, we shall first discretize continuum field theories on spatial lattices, providing a very elementary and accessible analysis of SSB. We will then turn to graphs and networks, and the properties of SSB on such geometric backgrounds.


\section{Lattices}

\subsection{Shift symmetry and Goldstone bosons}

Our practical discussion will evolve around a very simple free field theory in $d$ spatial dimensions defined by the action
\beq\label{freephi}
S=\frac12\int dt\,d\vec{x}\,\Big[(\del_t\phi)^2-(\nabla \phi)^2\Big]
\eeq
for the real scalar field $\phi(t,\vec{x})$, and the relevant symmetry whose spontaneous breaking we shall be discussing will be the very simple global shift symmetry
$\phi\to\phi+a$ respected by the above action.

At the first sight, the choice of this elementary setup may seem like an oversimplification, but that is not so. The shift symmetry is a prototype of many more sophisticated cases of SSB of continuous symmetries, while the field $\phi$ is nothing but the celebrated Goldstone boson. Indeed, consider the textbook example
of SSB for the $U(1)$ phase rotation symmetry of a complex scalar field $h$ with a `Mexican hat' potential:
\beq
S_h=\frac12\int dt\,d\vec{x}\,\Big[|\del_t h|^2-|\nabla h|^2-\lambda(|h|^2-h_0^2)^2\Big].
\eeq
Introducing the parametrization $h=(h_0+\eta)e^{i\phi}$, we obtain the following action for the real scalar fields $\eta$ and $\phi$:
\beq
S_h=\frac12\int dt\,d\vec{x}\,\Big[h_0^2(\del_t\phi)^2-h_0^2(\nabla \phi)^2+(\del_t\eta)^2-(\nabla \eta)^2-4\lambda h_0^2\eta^2+\mathrm{interactions}\Big].
\eeq
The two fields are thus decoupled at linear order. The field $\eta$ is massive, and its vacuum state is localized around the field configuration $\eta=0$ (this will be visible, in particular, from our subsequent considerations). The action for the field $\phi$ is exactly the same as (\ref{freephi}), up to normalization. The key question of SSB is whether the ground state of $\phi$ is localized around some particular field value, in which case the field-shift symmetry is broken, or if it is not, the symmetry is `restored.' Note that the field-shift symmetry of $\phi$ is exactly identical to the original $U(1)$ phase rotation symmetry of $h$, that is, $h\to e^{ia} h$. The field $\phi$ is nothing but the Goldstone boson associated with SSB, while its shift symmetry is a representation of the original continuous symmetry whose SSB one is trying to examine. (One often hears about the absence or presence of Goldstone bosons as such, but what is meant by that is whether the Goldstone field has conventional properties with a well-defined expectation value and particle excitations, as happens in cases with SSB \cite{MaRajaraman}.)

We shall therefore turn for the remainder of the exposition to free actions with field-shift symmetries of the form (\ref{freephi}) and their discretizations and generalizations.

\subsection{Discretization}

Quantization of the action (\ref{freephi}) leads, of course, to a completely elementary free field theory. But even in this setting, one would have to deal with the `ultraviolet divergences' coming from modes of arbitrarily short wavelength. For example, the variance of $\phi(t,\vec{x})$ at a single spatial point would turn out to be infinite in the ground state. This can be naturally dealt with by smearing the fields over spatial regions, but from the perspective of this essay, it is much more natural to discretize the continuous space described by $\vec{x}$ instead.

We thus replace each coordinate $x_n$ with $n=1,2,\cdots,d$ with a discrete set of values
\beq
x_n(i_n)=a i_n,\qquad i_n\in Z,
\eeq
forming a hypercubic lattice with lattice spacing $a$. We then discretize the action (\ref{freephi}) using the standard lattice representation for spatial derivatives, which can be written compactly as
\beq\label{freedscra}
S=\frac12\int dt\,\Big[\sum_I(\del_t\phi_I)^2-a^{-2}\sum_{\langle IJ\rangle}(\phi_I-\phi_J)^2\Big],
\eeq
where we have introduced the collective index $I$ labelling all points of the lattice, $I\equiv\{i_1i_2\cdots i_d\}$, while $\langle IJ\rangle$ denotes summation over all unordered pairs $(I,J)$ representing nearest neighbor sites, that is, the strings $\{i_1i_2\cdots i_d\}$ and $\{j_1j_2\cdots j_d\}$ must differ in only one position, and in that position, by exactly one unit. The collective indices have been introduced to facilitate transition from lattices to more general discrete structures in the second half of this essay.

By trivial rescaling of $t$ and $S$, we can eliminate $a$ altogether (it would only have to be reintroduced if we want to return to the continuum limit) to obtain the following simple discretization of (\ref{freephi}):
\beq\label{freedscr}
S=\frac12\int dt\,\Big[\sum_I(\del_t\phi_I)^2-\sum_{\langle IJ\rangle}(\phi_I-\phi_J)^2\Big]=\frac12\int dt\,\Big[\sum_I(\del_t\phi_I)^2-\sum_{IJ}\phi_IL_{IJ}\phi_J\Big].
\eeq
In the last representation, we introduced the {\it lattice Laplacian} $L_{IJ}$ defined as
\beq\label{latticeL}
L_{IJ}\equiv \begin{cases}
\,2d&\mbox{if }I=J,\\
-1&\mbox{if }I\mbox{ is a neighbor of }J.
\end{cases}
\eeq
(Note that $L$ provides a discretization of $-\nabla^2$, and it is nonnegative-definite.)
We are thus left with a system of coupled oscillators in (\ref{freedscr}). It should be easy to quantize it and study the ground state.

We note that discretizations like (\ref{freedscr}) are not of the form most typically used: only the space is discretized, leaving behind a mechanical system.
Similar structures have been explored on occasions, however---perhaps most prominently in the Hamiltonian lattice formulation of gauge theories \cite{Hamiltonian}. 
Lattices of coupled oscillators are also familiar from discussions of crystal vibrations and phonons \cite{Kittel}.

Canonical quantization of (\ref{freedscr}) amounts to writing the Hamiltonian
\beq\label{Hphi}
H=\frac12\sum_I\pi_I^2+\frac12\sum_{IJ}\phi_IL_{IJ}\phi_J
\eeq
and imposing the commutation relation
\beq\label{piphicomm}
[\pi_I,\phi_J]=-i\de_{IJ}
\eeq
between the operators $\phi_I$ and their conjugate momenta $\pi_I$. The equations of motion
\beq\label{piphiEOM}
\del_t\pi_I=-\sum_JL_{IJ} \phi_J,\qquad \del_t\phi_I=\pi_I
\eeq
are solved by
\beq\label{piphisol}
\phi_I=\sum_Q \frac1{\sqrt{2\om_Q}} u_{Q,I} (a^{\phantom{\dagger}}_Q e^{-i\om_Qt}+\ad_Q e^{i\om_Qt}),\quad \pi_I=-i\sum_Q \sqrt{\frac{\om_Q}2} \,u_{Q,I}(a^{\phantom{\dagger}}_Q e^{-i\om_Qt}-\ad_Q e^{i\om_Qt}).
\eeq
Here, $u_{Q,I}$ are the normalized mode function given by the eigenvectors of $L$ with eigenvalues $\lambda_Q$, so that
\beq
\sum_J L_{IJ}u_{Q,J}=\la_Qu_{Q,I},\qquad \sum_I u_{Q,I} u_{Q',I}=\de_{QQ'},\qquad \om_Q^2\equiv\la_Q,
\eeq
while  $\ad_Q$ and $a_Q$ are creation-annihilation operators satisfying
\beq
[a^{\phantom{\dagger}}_Q,\ad_{Q'}]=\de_{QQ'},
\eeq
as follows from (\ref{piphicomm}). Then, the ground state $|0\rangle$ of (\ref{Hphi}), which is our main interest here, is defined by being annihilated by all $a_Q$:
\beq
\forall Q:\qquad a_Q|0\rangle=0.
\eeq

\subsection{The ground state and its fluctuations}\label{secground}

With the very simple quantization of the free theory (\ref{freedscr}), it remains to examine its ground state and judge whether the fluctuations are small enough to keep
the system localized in the field space, so that the field shift symmetry is spontaneously broken. To keep the derivations completely clean, we will consider a finite hypercubic lattice, periodically identified, with $\ell$ lattice units in each direction, so that the total number of sites is 
\beq
V=\ell^d. 
\eeq
The relevant question is, of course,
how the fluctuations behave in the limit $\ell\to\infty$.

Our main point of interest is the fluctuations of the field values $\phi_I(0)$ at a given site within the vacuum state $|0\rangle$. These are best explored by considering the probability distribution function for $\phi_I(0)$ to equal $\ph$, which can be computed as
\beq\label{pphdef}
p_I(\ph)=\langle 0|\de[\phi_I(0)-\ph]|0\rangle=\frac1{2\pi}\int dk \,e^{-ik\ph}  \langle 0|e^{ik\phi_I(0)}|0\rangle.
\eeq
To compute the expectation values, we need to restore a small subtlety that has been overlooked in (\ref{piphisol}). Namely, due to the evident property
\beq
\sum_J L_{IJ}=0,
\eeq 
the lattice Laplacian $L$ always admits a zero mode described by the vector $(1,1,\cdots,1)/\sqrt{V}$, and this mode moves (in the field space) like a free quantum particle and not like a harmonic oscillator. We can assign the label $Q=0$ to this mode. The remaining modes with $Q=1,2,\cdots,V-1$ are described exactly as in (\ref{piphisol}). To incorporate this accurately in (\ref{piphisol}), we write
\beq\label{phizeromode}
\begin{split}
&\phi_I=\frac{X+P t}{\sqrt{V}}+\sum_{Q=1}^{V-1} \frac1{\sqrt{2\om_Q}} u_{Q,I} (a^{\phantom{\dagger}}_Q e^{-i\om_Qt}+\ad_Q e^{i\om_Qt}),\\
&\pi_I=\frac{P}{\sqrt{V}}-i\sum_{Q=1}^{V-1} \sqrt{\frac{\om_Q}2} u_{Q,I} (a^{\phantom{\dagger}}_Q e^{-i\om_Qt}-\ad_Q e^{i\om_Qt}).
\end{split}
\eeq
where $P$ and $X$ with $[P,X]=-i$ are the position and momentum operators of the zero-mode (representing a free quantum particle moving in the field space), and they commute with $a_Q$. With the zero mode properly taken into account, we have the following orthogonality and completeness relations, and decomposition of $L$:
\beq
\sum_I u_{Q,I}=0,\quad \sum_I u_{Q,I}u_{Q',I}=\de_{QQ'},\quad\sum_{Q=1}^{V-1}u_{Q,I}u_{Q,J}+\frac1V=\de_{IJ},\quad \sum_{Q=1}^{V-1}\om_Q^2u_{Q,I}u_{Q,J}=L_{IJ}.
\eeq
With this, one easily checks that (\ref{phizeromode}) satisfy the Heisenberg equations of motion (\ref{piphiEOM}) and the commutation relations (\ref{piphicomm}).

The vacuum state $|0\rangle$ is defined by $a_Q|0\rangle=0$ for $Q=1..V-1$, but for the zero mode, there is no normalizable ground state (the lowest energy state is annihilated by the momentum operator $P$ and is not normalizable). What we should consider physically is a normalized wavepacket whose wavefunction is $\Psi_0(X)$. With the normalization of $X$ chosen in our definition, any such normalizable $\Psi_0$ is irrelevant for large systems $V\to\infty$, and will reduce in this limit to a particle at rest at the origin (as the effective mass $V$ diverges and the particle becomes classical). Even though being at the origin (in the field space) looks like there is a preferred point and the field shift symmetry is spontaneously broken, it can be restored by the fluctuations coming from $a_Q$, which is the main effect we are considering here.

With these more accurate specifications of $|0\rangle$, we can compute (\ref{pphdef}) as
\begin{align}
p_I(\ph)=&\langle 0|\de[\phi_I(0)-\ph]|0\rangle=\frac1{2\pi}\int dk \,e^{-ik\ph}\left[  \int d\xi e^{-ik\xi/\sqrt{V}}\Psi_0(\xi)\right]\\
&\hspace{5cm}\times\prod_{Q=1}^{V-1} \langle 0|\exp\big[ik(2\om_Q)^{-1/2} u_{Q,I} (a^{\phantom{\dagger}}_Q +\ad_Q)]\big|0\rangle.\nn
\end{align}
The first line is merely a Fourier transform of $\Psi_0$, that is, $\tilde\Psi_0(q)\equiv\int d\xi \,e^{-iq\xi}\Psi_0(\xi)$, while the second line is computed by the Baker-Campbell-Hausdorff formula according to
\beq
\langle 0|e^{C(a+\ad)}|0\rangle=\langle 0|e^{C\ad}e^{C a}e^{-C^2[\ad,a]/2}|0\rangle=e^{C^2/2}.
\eeq
Hence,
\beq
p_I(\ph)=\frac1{2\pi}\int dk \,e^{-ik\ph}\,\tilde\Psi_0(k/\sqrt{V})\exp\left[-\frac{k^2}4\sum_{Q=1}^{V-1}\frac1{\om_Q}u_{Q,I}^2\right].
\eeq
At $V\to\infty$, $\tilde\Psi_0$ yields an irrelevant $k$-independent constant, and the remaining Gaussian integral yields
\beq\label{dstrW}
p_I(\ph)\sim e^{-\phi^2/W_I^2},\qquad W_I^2\equiv \sum_{Q=1}^{V-1}\frac1{\om_Q}u_{Q,I}^2.
\eeq
The distribution of field values is thus Gaussian and centered at $\ph=0$. If $W$ is finite, this spontaneously breaks the field shift symmetry $\ph\to\ph+a$. But the central point of the considerations in the essay is that as the system becomes infinitely large with $V$ going to $\infty$, $W$ does not have to stay finite. If $W$ diverges, the ground state is delocalized in the field space, and there is no SSB. More than that, as $W$ tends to infinity, the Gaussian distribution of width $W$ becomes completely uniform, making sure that no preferred value of the field may be indicated, and the field shift symmetry is restored. Our main technical question is then: {\it under what conditions does $W$ diverge at $V\to\infty$?}

We can simplify the expression for $W$. Due to the perfect spatial translation symmetry of the hypercubic lattice, $W_I$ cannot depend on the vertex number $I$. We can then tautologically sum it over all vertices and divide by $V$, without changing its value:
\beq\label{WVQ}
W_I^2=W^2\equiv \frac1V\sum_I \sum_{Q=1}^{V-1}\frac1{\om_Q}u_{Q,I}^2=\frac1V\sum_{Q=1}^{V-1}\frac1{\om_Q}.
\eeq
All we need to know to judge the presence or absence of SSB is the spectrum of the lattice Laplacian $\la_Q\equiv \om_Q^2$ at large $V$.

The spectra of lattice Laplacians are the familiar phonon spectra of condensed matter physics. To compute them, we decode the collective indices $I$ and $J$ in the eigenvalue equation $\sum_J L_{IJ}u_J=\la u_I$ as $\{i_1\cdots i_d\}$ and $\{j_1\cdots j_d\}$ and take into account the definition of $L$ in (\ref{latticeL}) to obtain
$$
2d\,u_{i_1\cdots i_d}-u_{i_1+1,i_2\cdots i_d}-u_{i_1-1,i_2\cdots i_d}-u_{i_1,i_2+1\cdots i_d}-u_{i_1,i_2-1\cdots i_d}-\ldots-u_{i_1,i_2\cdots i_d+1}-u_{i_1,i_2\cdots i_d-1}=\la \,u_{i_1\cdots i_d}.
$$
This is solved by discretized standing waves
\beq
u_{i_1\cdots i_d}=e^{i\sum_{s=1}^d q_si_s},\qquad \la=4\sum_{s=1}^d\sin^2 \frac{q_s}2,
\eeq
where $q$ must be of the form
\beq
q_s=\frac{2\pi l_s}{\ell},\qquad l_s=0,1,\cdots,\ell-1,
\eeq
in order to satisfy the periodic boundary conditions of the $\ell\times\ell\times\cdots\times\ell$ lattice.

We then have to estimate, at large $V\equiv \ell^d$,
\beq\label{Wsin}
W^2=\frac1{\ell^d}{\sum_{l_s=0}^{\ell-1}}\rule{0mm}{4mm}^{\hspace{-0.2mm}\smallcirc{0.7}}\,\frac1{\om_Q}=\frac1{2\ell^d}{\sum_{l_s=-\lfloor\ell/2\rfloor}^{\lceil\ell/2\rceil-1}}\rule{0mm}{4mm}^{\hspace{-4mm} \smallcirc{0.7}}\,\left(\sum_{s=1}^d\sin^2 \frac{\pi l_s}{\ell}\right)^{-1/2},
\eeq
where the empty circles attached to the sum signs indicate that the point $l_s=0$ (for all $s$) must be excluded.  While it is possible to make shortcuts for judging the behavior of this expression at $\ell\to\infty$, as we shall see below, it is instructive, for once, to process this expression by brute force in the purely discrete-sum notation and extract its divergence.

To do so, consider first a set of numbers $y_s$ such that $|y_s|\le\pi/2$.
Then,
\begin{align*}
&\left|\left(\sum_{s=1}^d\sin^2 y_s\right)^{-1/2}-\left(\sum_{s=1}^d y^2_s\right)^{-1/2}\right|\,\,=\,\,\left(\sum_{s=1}^d\sin^2 y_s\right)^{-1/2}-\left(\sum_{s=1}^d y^2_s\right)^{-1/2}\\
&=\frac{\sum_s(y^2_s-\sin^2 y_s)}{\big[(\sum_s\sin^2 y_s)^{1/2}+(\sum_s y^2_s)^{1/2}\big]\sqrt{(\sum_s\sin^2 y_s)(\sum_s y^2_s)}}\le\frac{\sum_s(y^2_s-\sin^2 y_s)}{2(\sum_s\sin^2 y_s)^{3/2}},
\end{align*}
where we have repeatedly used $|y|\ge| \sin y|$. The last expression is finite and positive: whenever there is at least one nonzero $y_s$, it is manifestly finite, so the only subtlety is its behavior when all $y_s$ tend to 0. But then, if we write $y_s=\eta_s y$ and send $y$ to 0, the numerator is $\sim y^4$ and the denominator is $\sim y^3$, so there is no divergence. Because of this estimate, we can replace $\sin^2\pi\l_s/\ell$ in (\ref{Wsin}) with $(\pi\l_s/\ell)^2$ without influencing the part of $W$ divergent at $\ell\to\infty$. Indeed, we are making a mistake in the summand  that is finite at $\ell\to\infty$, and since there are $\ell^d$ summation terms, the result of summation is at most of order $\ell^d$, so it cannot give a divergent contribution at large $\ell$ after being multiplied with $1/\ell^d$. Hence, the divergence of $W$ at large $\ell$ can be judged from
\beq\label{Wsum}
W^2\sim\frac1{2\ell^d}{\sum_{l_s=-\lfloor\ell/2\rfloor}^{\lceil\ell/2\rceil-1}}\rule{0mm}{4mm}^{\hspace{-4mm} \smallcirc{0.7}}\,\left(\sum_{s=1}^d\frac{\pi^2 l_s^2}{\ell^2}\right)^{-1/2}=\frac{\ell^{1-d}}{2\pi}{\sum_{l_s=-\lfloor\ell/2\rfloor}^{\lceil\ell/2\rceil-1}}\rule{0mm}{4mm}^{\hspace{-4mm} \smallcirc{0.7}}\,\left(\sum_{s=1}^d l_s^2\right)^{-1/2}.
\eeq
If $d=1$, this expression diverges as $\log\ell$. Indeed, we get up to constant factors
\beq
W^2\sim{\sum_{l=-\lfloor\ell/2\rfloor}^{\lceil\ell/2\rceil-1}}\rule{0mm}{4mm}^{\hspace{-3mm} \smallcirc{0.7}}\,\,\,\frac1{l}\ge 2\sum_{l=1}^{\lceil\ell/2\rceil-1}\frac1{l}=2\int_1^{\lceil\ell/2\rceil}\frac{dx}{\lfloor x\rfloor}>2\int_1^{\lceil\ell/2\rceil}\frac{dx}{x}\sim\log\ell.
\eeq
If $d>1$, $W$ receives no divergent contributions. Indeed, the sum over $l_s$ will diverge as $\ell^{d-1}$, which will be exactly compensated by the explicit factor $\ell^{1-d}$. To see that the sum over $l_s$ diverges (at most) as $\ell^{d-1}$, observe that the summand is positive so the expression only grows if we extend the summation region. So we can bound $W$ from above by extending the sum over $l_s$ to all lattice points at distance not less than 1 and not more than $\lceil\ell/2\rceil\sqrt{d}$ from the origin, since this includes all points appearing in the original summation in (\ref{Wsum}). We then introduce the function $n(r)$ counting the number of lattice points whose distance from the origin is in $[r,r+1)$. We can then upper-bound $W$ at large $\ell$ by (up to constant factors)
$$
\ell^{1-d}\sum_{r=1}^{\lceil\ell/2\rceil\sqrt{d}}\frac{n(r)}{r}.
$$
But $n(r)$ grows as $r^{d-1}$ at large $r$ \cite{latticecount}, so one has a sum of $r^{d-2}$ over $r$ that will diverge as $\ell^{d-1}$ at large $\ell$ by the usual Faulhaber's formula. As a result, $W$ stays finite.

We have thus reproduced the classic result of Coleman \cite{Coleman} on the absence of SSB in one spatial dimension, but without any `ultraviolet' divergence issues associated with standard QFT computations. All expressions are finite at fixed $\ell$, while the localization range of fields diverges as $\ell$ increases in one dimension, but not in higher dimensions. As a result, there is no preferred value in the field space in 1d, but in higher dimensions, the configurations contributing to the ground state are localized around $\phi=0$, with finite fluctuations. Field shifts would of course produce equivalent configurations localized around other values of $\phi$, but all of them will have finite fluctuations with a measurable preferred reference value in the field space. This is a physical manifestation of the SSB of the field shift symmetry.

Intuitively, large field fluctuation in a 1d chain occur because the chain is not well-connected (for instance, it can be cut by removing one site). If we want to have different field values at two distinct site, the transition may happen in a finite localized region between the two sites, and there will be a finite energy cost. This allows the field values to drift easily (especially if they drift slowly) as one moves along the chain. But in higher dimensions, the interface between two regions with different field values is itself a higher-dimensional surface (or a line in 2d), and this yields bigger energy costs, limiting how much fields can fluctuate.

The divergence in (\ref{WVQ}) corresponding to a 1d chain could be understood in a different way, without doing the detailed computations above. Indeed, (\ref{WVQ}) can be equivalently rewritten as
\beq\label{Wrho}
W^2=\int_{0}^{\infty} \frac{d\la}{\sqrt{\la}}\,\rho(\la),
\eeq
where $\rho(\la)$ is the eigenvalue density of the lattice Laplacian, defined as 
\beq
\rho(\la)\equiv \frac1V\sum_Q \de(\la-\la_Q).
\eeq
As the number of vertices $V$ goes to infinity, $\rho$ usually converges to a continuous curve (in the very least, it is a good probability measure). Then, the question of whether $W$ is finite amounts to asking whether the integral in (\ref{Wrho}) is convergent at $\la=0$. This evidently happens if the blow-up of $\rho(\la)$ at $\la=0$ is weaker than $1/\sqrt{\la}$.

Singularities in spectral densities are commonly studied in condensed matter physics under the name of Van Hove singularities \cite{Kittel}. As is well-known, the asymptotics of $\rho(\la)$ at $\la=0$ in $d$ dimensions is $\la^{d/2-1}$. This automatically recovers the divergence we found in $W$ in 1d. We will return to similar formulas for the asymptotics of the spectral density in our discussions of the spectral dimension in the next section. (More exotic higher-order Van Hove singularities have been actively discussed in recent years \cite{HOVHS}, but seem to refer to the singularities in higher bands rather than at the bottom of the lowest acoustic band, which is of interest for us here.)

Before proceeding with generalizations, it is wise to dwell for a moment on the particularly interesting and natural implementation of field shift symmetries in 
worldvolume theories of extended solitonic objects. Classical topological solutions with energies localized around lower-dimensional submanifolds (solitons, vortex lines, domain walls) are known to translate into dynamical objects in the corresponding quantum theory \cite{Rajaraman}. The transverse fluctuations of such objects (for example, vibrations of a quantum vortex line) are described by lower-dimensional massless scalar QFTs living on the worldvolume, with the fields corresponding to the local displacement of the solitonic object in the transverse directions. The field shift symmetry literally becomes the translational symmetry in the transverse direction. Because there is no SSB in 1d, this implies that quantum vortex lines behave differently from higher-dimensional extended solitons \cite{thesis,localr,EvslinLiu}. In particular, transverse momentum is well-defined and conserved, and the final transverse momentum of the vortex line following a scattering event must reflect that. Failure to account for this effect while expanding around classical solutions corresponding to static vortex lines leads to infrared divergences at higher orders. This is analogous to the much better known recoil effect \cite{thesis,mesonsoli,D0r,EvslinGuo,SenStefanski} for fully localized point-like solitons. Evidently, in that case, the `worldline' theory describing the transverse fluctuations is quantum mechanics where SSB is impossible as there are finitely many degrees of freedom. That is why transverse momentum is conserved, and the soliton starts moving (recoils) due to the impact of fundamental quanta in what is known as a meson-soliton scattering process.

\subsection{Nonrelativistic QFTs}

While we started with a relativistic field theory (\ref{freephi}), the discretization broke the symmetry between space and time. Thereafter, we are in a format much closer to nonrelativistic field theories (though the lattice reduces the symmetry even further), and it is natural to include such systems into consideration.

Before turning to nonrelativistic fields, it is useful to revisit the setup described above for a moment, but now with the scalar field mass included---indeed, quanta of massive fields connect naturally to nonrelativistic kinematics. This amounts to adding $-m^2\phi^2/2$ to the action (\ref{freephi}), which will naturally propagate though the formulas, replacing the Laplacian $L$ with $L+m^2I$, where $I$ is the identity matrix.
With this replacement, (\ref{WVQ}) turns into
\beq
W^2(m)=\frac1V\sum_{Q=1}^{V-1}\frac1{\sqrt{\om_Q^2+m^2}}=\int_0^{\infty}\frac{d\lambda}{\sqrt{\lambda+m^2}}\,\rho(\lambda),
\eeq
with the Laplacian eigenvalue density $\rho(\lambda)$.
But this expression can never diverge due to contributions near $\lambda=0$. Whatever singularity $\rho(\lambda)$ may have near the origin,
this singularity must be integrable due to the probability normalization of $\rho(\lambda)$. On the other hand, $1/\sqrt{\lambda+m^2}$ approaches a constant at the origin and cannot make this singularity any stronger, so that the integral is convergent. Unsurprisingly, massive fields are always localized near the value $\phi=0$ in the field space.

This situation extends to nonrelativistic Schr\"odinger fields obtained from complex-valued massive fields in the low-energy limit. Such fields arise as Type B Goldstone bosons when breaking noncommuting symmetries in nonrelativistic systems \cite{Brauner, WB, Kapustin, WM}, and they are described, in appropriate units, by the action
\beq\label{freeSch}
S=\int dt\,d\vec{x}\,\Big[i\big(\bar\psi\del_t\psi-\psi\del_t\bar \psi\big)-|\nabla \psi|^2\Big].
\eeq
This is discretized to
\beq\label{freeSchdscr}
S=\int dt\,\Big[i\sum_I\big(\bar\psi_I\del_t\psi_I-\psi_I\del_t\bar \psi_I\big)-\sum_{IJ}\bar\psi_I L_{IJ}\psi_J\Big],
\eeq
with the shift symmetry $\psi_I\to\psi_I+a$.
Upon quantization, one obtains operators $\psi$ and $\psi^\dagger$, satisfying
\beq
i\,\del_t\psi_I=\sum_J L_{IJ}\psi_J, \qquad [\psi_I,\psi^\dagger_J]=\de_{IJ},
\eeq
solved in analogy to (\ref{phizeromode}) by
\beq
\psi_I=\frac{X+iP}{\sqrt{2V}}+\sum_{Q=1}^{V-1}a_Q u_{Q,I}e^{-i\om_Qt},\qquad [a_Q,a^\dagger_{Q'}]=\de_{QQ'}-\frac1{V},\qquad [X,P]=i.
\eeq
The key difference with (\ref{phizeromode}) is that, due to the first-order equations of motion, the factor $\sqrt{\om_Q}$ in the denominator is absent.
As a result, $1/\sqrt{\lambda}$ will be absent in (\ref{Wrho}), and the integral will be convergent, yielding a finite $W$. Thus, a quantum Schr\"odinger field in its ground state is always localized in field space, just like a massive relativistic field, and the field shift symmetry is always spontaneously broken.

\subsection{Higher-derivative theories}\label{sechigher}

Once the Lorentz invariance is relaxed to nonrelativistic symmetries, and there are no longer symmetry transformations relating space and time,
the number of spatial derivatives in the equations of motion is no longer tied to the number of time derivatives. It is thus natural to consider higher-derivative generalizations of (\ref{freephi}), which have been explored comprehensively in \cite{GGHY}. In such theories, the shift symmetry modifying the field by a position-independent constant is enhanced to shifts by arbitrary polynomials of the coordinates (with the permitted degrees of polynomials constrained by the number of spatial derivatives in the action). The prototypical equations of motion of this sort are
\beq
\del_t^2\phi+(-\nabla^2)^p \phi=0
\eeq
that reduce at $p=1$ to the ordinary wave equation for (\ref{freephi}). While such equations may appear awkward from a high-energy physics standpoint,
they are in fact perfectly physical, and the case $p=2$ in two dimensions describes linearized transverse vibrations of solid plates \cite{plates}.

In the latticized version of field theories considered here, a natural analog of such equations is 
\beq
\del_t\phi_I+\sum_J [L^p]_{IJ}\phi_J=0,
\eeq
involving the $p\hspace{0.3mm}$th power of the lattice Laplacian (\ref{latticeL}). As these equations are structurally identical to the ones considered above, except that $L$ gets replaced by $L^p$, all the derivations will go through, except that in the solutions of quantum equations of motion given by (\ref{phizeromode}), one will have
\beq
\om_Q^2=\lambda^p_Q,
\eeq
where $\lambda_Q$ are the nonzero eigenvalues of $L$ as before. Then, instead of (\ref{Wrho}), one will have for the localization range of the ground state in the field space
\beq
W^2=\int_{0}^{\infty} \frac{d\la}{\la^{p/2}}\,\rho(\la).
\eeq
As discussed above, for a hypercubic lattice in $d$ dimensions, $\rho(\la)$ behaves as $\la^{d/2-1}$ near $\la=0$. Hence, $W$ will be divergent if
\beq\label{divgenlat}
\frac{p-d}2+1\ge 1\quad \to\quad p\ge d.
\eeq
For $p=1$, this reduces to the (discretized form of the) classic result of Coleman in one dimension, while the higher $p$ case agrees with the continuum version in \cite{GGHY}. The key message is that for higher-derivative field theories, the absence of spontaneous symmetry breaking persists to higher dimensions than it does for conventional relativistic fields.


\section{Graphs}

\subsection{Quantum fields on general graphs and graph Laplacians}

The choice of notation in the previous section may have appeared unconventional to some readers since, in particular, the lattice sites were labeled by a single integer collective index, rather than by the usual $d$-tuple of integers representing discretized coordinates. An advantage of this notation is that it translates without change into a description of general discrete geometries represented by graphs, as we shall see immediately, and while every lattice is a graph, not every graph is a lattice.

Indeed, on a general graph with $V$ vertices, we could restate formula (\ref{freedscr}) verbatim:
\beq\label{freegrph}
S=\frac12\int dt\,\Big[\sum_I(\del_t\phi_I)^2-\sum_{\langle IJ\rangle}(\phi_I-\phi_J)^2\Big]=\frac12\int dt\,\Big[\sum_I(\del_t\phi_I)^2-\sum_{IJ}\phi_IL_{IJ}\phi_J\Big],
\eeq
except that, now, $\phi_I$ is the value of the field on the graph vertex number $I$ and $\langle IJ\rangle$ means summing over all pairs of vertices $I$ and $J$ connected by a graph edge. The graph Laplacian $L$ can be defined by the quadratic form in the above expression, while it can also be expressed directly through the adjacency matrix $A_{IJ}$, which is a symmetric zero-diagonal matrix whose entry $A_{IJ}$ equals 1 if there is an edge connecting vertices $I$ and $J$, and zero otherwise. Given this adjacency matrix, we can define the vertex degrees $d_I\equiv \sum_J A_{IJ}$, and then
\beq\label{graphL}
L_{IJ}=D_{IJ}-A_{IJ},\qquad D=\mathrm{diag}(d_1,d_2,\cdots,d_V),
\eeq
which agrees with the above definition in terms of quadratic forms. Without loss of generality, we assume the graph to be connected, as different disconnected components would have to be treated completely independently to assess the presence of SSB within each component.

Of course, there is no unique way to generalize Laplacians from a lattice (which is a regular graph) to graphs with heterogeneous degrees, since one may insert, for example, arbitrary degree-dependent modifications, which would reduce to trivial rescalings when all degrees are the same. We could thus write, instead of (\ref{freegrph}),
\beq\label{freegrphmod}
S=\frac12\int dt\,\Big[\sum_I(\del_t\phi_I)^2-\sum_{\langle IJ\rangle}w(d_I,d_J)(\phi_I-\phi_J)^2\Big],
\eeq
with an arbitrary symmetric function $w$. The particular choice $w(d,d')=1/\sqrt{dd'}$ corresponds to the so-called {\it normalized} graph Laplacian, used instead of the ordinary (combinatorial) graph Laplacian (\ref{freegrph}) in some applications. We shall mostly focus, however, on the particularly simple generalization of the conventional lattice discretization
of continuous field theories provided by (\ref{freegrph}). Graph Laplacians have found countless applications in network science, including their recent prominent use for
network renormalization \cite{laprenorm}.

The graph Laplacian (\ref{graphL}) has the same formal properties as the lattice Laplacian from the previous section: it has one zero eigenvalue (assuming that the graph is connected) corresponding to the all-one vector, and $V-1$ positive eigenvalues. We can repeat the derivations of section~\ref{secground} verbatim, obtaining an identical solution for free quantum fields (\ref{phizeromode}) and for their onsite variance (\ref{WVQ}). It remains to understand how these expressions depend on the actual geometry of the underlying graph, for which many more options exist than what is available in the setting of simple hypercubic lattices.

In what follows, we work with the `higher-derivative' generalization of ordinary free fields, as described in section~\ref{sechigher}, now transplanted to graphs. The corresponding action is
\beq\label{freegrphhigher}
S=\frac12\int dt\,\Big[\sum_I(\del_t\phi_I)^2-\sum_{IJ}\phi_I(L^p)_{IJ}\phi_J\Big].
\eeq
The derivations remain completely unchanged structurally, except that, in (\ref{WVQ}), one must replace $\om_Q^2=\lambda_Q$ with $\om_Q^2=\lambda_Q^p$.

\subsection{Fractional resistance distance, Kirchhoff index and spectral dimension}

On a general graph, there is no counterpart of translation symmetry, and the vertices are in general inequivalent. For that reason, $W_I$ defined by (\ref{dstrW}) depends, in general, on the vertex number $I$. One can even imagine that $W_I$ would diverge on some sites and not on others as the graph grows, though admittedly, this situation is somewhat exotic. (As a trivial example, consider an $\ell\times \ell$ two-dimensional square lattice with the endpoints of a one-dimensional lattice of length $O(\ell)$ or more attached to two of its sites. The symmetry will be broken on the two-dimensional component, but restored on the long one-dimensional loop as $\ell$ tends to $\infty$.)

The most detailed analysis then consists in examining $W_I$ vertex-by-vertex:
\beq\label{Wgrph}
 W_I^2=\sum_{Q=1}^{V-1}\frac1{\om_Q}u_{Q,I}^2=\sum_{Q=1}^{V-1}\frac1{\lambda_Q^{p/2}}\,u_{Q,I}^2=(L^{-p/2})_{II}.
\eeq
Here, $L^{-\gamma}$ is understood as $(L^{-1})^\gamma$, where $L^{-1}$ is the {\it pseudoinverse} of $L$ defined as
\beq
L^{-1}_{IJ}=\sum_{Q=1}^{V-1}\frac1{\lambda_Q}\,u_{Q,I}u_{Q,J},
\eeq
so that the zero eigenvalue of $L$, corresponding to $Q=0$ and the all-one eigenvector, is ignored when the inversion is implemented.

The pseudoinverse of the graph Laplacian appears prominently in the construction of {\it resistance distances} on the graph \cite{KleinRandic}, which are defined as the electrical resistance between vertices $I$ and $J$ if each edge is interpreted as an ordinary conductor of resistance 1. The actual expression for such resistance distance $\Omega_{IJ}$ is
\beq
\Omega_{IJ}\equiv L^{-1}_{II}+L^{-1}_{JJ}-2L^{-1}_{IJ}.
\eeq
The sum of this quantity over all distinct pairs of vertices is known as the Kirchhoff index $K$:
\beq
K\equiv\frac12\sum_{IJ}\Omega_{IJ}.
\eeq

To connect with (\ref{Wgrph}), it is natural to define the `fractional analogs' of the above quantities:
\beq
\Omega^{(\gamma)}_{IJ}\equiv L^{-\gamma}_{II}+L^{-\gamma}_{JJ}-2L^{-\gamma}_{IJ},\qquad K_\gamma=\frac12\sum_{IJ}\Omega^{(\gamma)}_{IJ}.
\eeq
As $L$ annihilates the all-one vector, this property is shared by $L^{-\gamma}$:
\beq
\sum_J (L^{-\gamma})_{IJ}=0.
\eeq
Hence,
\beq
\sum_J\Omega^{(\gamma)}_{IJ}=VL^{-\gamma}_{II}+\sum_JL^{-\gamma}_{JJ},
\qquad K_\gamma=V\sum_I L^{-\gamma}_{II},\qquad 
L^{-\gamma}_{II}=\frac1V\sum_J\Omega^{(\gamma)}_{IJ}-\frac{K_\gamma}{V^2}.
\eeq
Then, $W_I$ of (\ref{Wgrph}) is expressed through the average fractional resistance distance between the point of observation and all other vertices, and the fractional Kirchhoff index:
\beq
W^2_I=\frac1V\sum_J\Omega^{(p/2)}_{IJ}-\frac{K_{p/2}}{V^2}.
\eeq
The case $p=2$ corresponds to ordinary resistance distance and Kirchhoff index. (Discussion of properties of resistance distance on random graphs can be seen in \cite{resist}.) Intuitively, this formula means that the symmetry is restored on those vertices that are remote from the bulk of the graph, in the precise sense provided by the fractional resistance distance. This, in turn, implies a form of low connectivity of the graph, analogous to what happens on low-dimensional lattices.

While vertices are inequivalent on general graphs, they are often statistically equivalent in realistic networks, and in any case, an instructive quantity to consider is the averaged $W_I^2$:
\beq\label{Wav}
\overline{W^2}\equiv\frac1V\sum_I W_I^2=\frac1V\,\mathrm{Tr}\,[L^{-p/2}]=\frac{K_{p/2}}{V^2}.
\eeq
Evidently, if $W^2$ diverges at large $V$, at least some of the $W_I$ must diverge. Furthermore, they should diverge sufficiently strongly and on a sufficient fraction of sites so as not to get diluted by division with $V$ in the process of averaging.

Representation (\ref{Wav}) connects to formulas like (\ref{Wrho}) in terms of the Laplacian eigenvalue density $\rho(\lambda)$. Indeed,
\beq\label{Wavrho}
\overline{W^2}=\int_0^\infty\frac{d\lambda}{\lambda^{p/2}}\,\rho(\lambda).
\eeq
Since $\rho$ itself is normalized as a probability distribution and integrable $\int d\lambda\,\rho(\lambda)=1$, the question of whether (\ref{Wavrho}) diverges reduces
to the strength of the singularity of $\rho(\lambda)$ at the origin. (There are some parallels between this formula and considerations in terms of the spectral density of the graph Laplacian seen in \cite{LPS}, where thermal --- rather than quantum --- fields on the regular trees are considered.)

Imagine that the spectral density of the Laplacian $\rho(\lambda)$ behaves as a power law at the origin:
\beq\label{spectrald}
\rho(\lambda)\sim \lambda^{d_s/2-1}.
\eeq
In that case, the onsite field variance on average, given by (\ref{Wavrho}), is infinite if
\beq\label{ineqspectral}
p\ge d_s,
\eeq
just like in (\ref{divgenlat}). One can recognize the parameter $d_s$ as nothing but the {\it spectral dimension} of the graph, one of the standard measures that generalize
the ordinary topological dimension of Euclidean spaces. The spectral dimension originated in condensed matter literature \cite{Dhar,HughesShlesinger,AlexanderOrbach,CassiRegina,BurioniCassi} as a quantification of diffusion properties on discrete geometries.
It later migrated into a variety of fields, including discrete quantum gravity \cite{universe}. 
In contemporary network science, the spectral dimension plays a key role in a few hot topics:
dynamical propeties of synchronization within the Kuramoto model \cite{specdim1},
studies of the `network geometry with flavor' model \cite{specdim2} 
and its deterministic counterparts such as the Apollonian networks \cite{specdim3}.
Our discussion makes it clear that it is precisely the spectral dimension, or more precisely, the spectral dimension $d_s$ defined through the ordinary (combinatorial) Laplacian, that controls --- via the condition (\ref{ineqspectral}) --- the presence of absence of spontaneous symmetry breaking
in QFTs on graphs/networks defined by (\ref{freegrphhigher}). Spectral dimension has previously emerged in considerations of the Mermin-Wagner theorem for graphs \cite{ssb1,ssb2}, closely related to the (zero-temperature) QFT case treated here, though the precise accounting differs, as it does between the classic considerations of Coleman and Mermin-Wagner. Further examples of the role played by the spectral dimension for statistical systems can be seen in \cite{sph,infsph}.

Before proceeding to discuss the properties of spectral dimension on structured graphs, it is wise to review the simplest case of random graphs:
the Erd\H{o}s-R\'enyi graphs where pairs of vertices are connected independently with a given probability, or the configuration model where the degree distribution
is prescribed, but the graph is otherwise random. Simple settings like this do not provide, however, the power law asymptotics (\ref{spectrald}), and their effective dimension is effectively infinite, ascertaining spontaneous breaking of the field shift symmetry in all cases. 
(To be more precise, we should be talking about the giant connected component of such graphs, while finite disconnected components trivially cannot support SSB since
only a finite number of degrees of freedom can reside on such components.)
The spectral densities of graph Laplacians in such models have been
studied in a number of works \cite{RB,onthem,laplace}. The lucid exposition of \cite{onthem} shows, in particular, that in the absence of vertices of degree 2, there is a spectral gap, and the spectral density $\rho(\lambda)$ is exactly zero below a certain threshold. If there is a finite fraction of vertices of degree 2, the support of $\rho(\lambda)$ extends all the way to $\lambda=0$, but the asympotics in that region is such that $\rho(\lambda)$ decays faster than any power. All of these cases correspond to an infinite spectral dimension.

\subsection{Fractals}

As we are discussing a generalization of standard hypercubic lattices that yields noninteger dimensions, the subject of fractals naturally enters the game.

Classic regular fractals, for instance, the ones due to Sierpi\'nski \cite{Sierpinski}, are obtained by iterative processes such as removal of regions of a plane, or adding line segments. They can in particular be formulated as limits of applying a recursive process to a graph, where a certain motif is repeatedly identified and replaced by a more elaborate motif including more vertices and edges, {\it ad infinitum}.

Studies of analytic properties of functions on such regular fractals, including the spectra and eigenfunctions of their Laplacians, have attracted considerable
attention from mathematicians in recent decades \cite{Strichartz, StrichartzTeplyaev}. For the simplest Sierpi\'nski gasket built by iterative insertion of triangle motifs,
the spectral dimension is well understood and equals $2\ln 3/ \ln 5 \approx 1.365$ \cite{HilferBlumen,Dunne}, which is comparable to but different from the more
familiar Hausdorff dimension $\ln 3/ \ln 2 \approx 1.585$. This spectral dimension will then determine which field theories possess ground states localized in the field space that break the shift symmetry, and which ones avoid SSB.

Another curious definition of recursive lattices goes back to the original work of Deepak Dhar \cite{Dhar} that spurred further studies of the spectral dimension.
There, the objective is precisely to construct a family of lattices whose spectral dimension can be tuned. The algorithm is to start with a complete graph on $q+1$ vertices. At each recursive step, each vertex---connected to $q$ neighbor vertices $\{V_{\mathrm{neighbor}}\}$---is replaced by a complete graph on $q$ vertices, with each of the new vertices additionally connected to exactly one vertex in $\{V_{\mathrm{neighbor}}\}$, simply inheriting the connections of the precursor vertex at the previous recursive step. The graph remains $q$-regular at all steps, while the spectral dimension equals (after infinitely many iterations)
\beq
d_s=2\,\frac{\ln q}{\ln (q+2)}.
\eeq
This expression varies between 1 and 2 as $q$ varies between 2 and $\infty$. Field theories on such lattices have been treated in \cite{Hill,nested}. Again, the phenomenology of spontaneous symmetry breaking will be controlled by the construction outlined above.

For lattices constructed recursively, properties of the eigenvalue density of their Laplacians are typically determined by a selfconsistent equation that can be arrived at 
via a number of different routes. There is a connection here to the selfconsistent equations used in studies of disorder on the Bethe lattice, see \cite{ACAT1,ACAT2} and the extensive follow-up. One particularly attractive technology in this regard is based on the representation of Laplacian resolvents in terms of supersymmetric integrals \cite{nuclphysb} (the random onsite potentials central to this line of work can be easily removed for our purposes, while the essential part is that the recursive structure to the graph gets effectively translated into a selfconsistent equation whose solution encodes the spectral density).

\subsection{Networks}

Regular fractal lattices are, in a sense, one step away from hypercubic lattices providing discretizations of ordinary Euclidean spaces. Yet, the general setting of graphs
supplies a vast range of opportunities for defining richer geometric structures.

The question of constructing network geometries with tunable spectral dimensions has been raised explicitly in the literature.
In \cite{tunable}, one starts with a simple 1d geometry: a ring of sites, each of them connected to its two neighbors to the left and to the right,
and then adds a sprinkling of extra links randomly in the spirit of `small-world' network models \cite{smallw}. More specifically, nodes $I$ and $J$ at distance $r_{IJ}$ from each other are connected with probability
\beq
p_{IJ}=\frac{1}{r_{IJ}^{1+\sigma}},
\eeq
where different notions of distance can be used successfully, see \cite{tunable} for details. The resulting spectral dimension is $\sigma$-dependent and 
interpolates between $1$ and $\infty$. (The discussion of \cite{tunable} uses a definition of the spectral dimension based on the normalized graph Laplacian, but the behavior for the combinatorial graph Laplacian that has been our main focus here is expected to be similar.)

An even more striking example is seen in \cite{heterogeneous}. There, networks with large degrees are considered, so there is no one-dimensional structure introduced at the start. Yet, power-law divergences in the Laplacian eigenvalue density are explicitly seen in some regimes, see Fig.~2 in that paper.
The power law in the spectral density directly reflects the power law in the degree distribution used to define the model, and it is therefore a tunable parameter allowing for adjusting the spectral dimension as one wishes.

In \cite{heterogeneous}, one finds a prominent discussion of networks with heterogeneous coupling strengths, where it is not only specified whether or not two nodes are connected, but there is also a prescription for the strength of this connection indicated by the `link weights' $w_{IJ}$. Such `link weights' provide for a natural generalization of (\ref{freegrph}) given by
\beq
S=\frac12\int dt\,\Big[\sum_I(\del_t\phi_I)^2-\sum_{\langle IJ\rangle}w_{IJ}(\phi_I-\phi_J)^2\Big].
\eeq
The tunable link strengths create extra options for the corresponding phenomenology. For example, disorder in the weights can induce phenomena similar to Anderson localization \cite{ACAT1,ACAT2,nuclphysb,molinari}. Even on an ordinary hypercubic lattice, if the disorder anisotropically affects only the weights in some of the directions, one can expect that effective propagation only spans the remaining directions, effectively reducing the dimensionality and strengthening the spectral singularities that in turn govern the absence of SSB.

Proceeding to further generalizations, one could make the field $\phi_I$ complex and replace $(\phi_I-\phi_J)^2$ with $|\phi_I-e^{i\theta_{IJ}}\phi_J|^2$, where
$\theta_{IJ}=-\theta_{JI}$. This ushers the discussion into the world of magnetic Laplacians \cite{magnetic}, as the phases $\theta_{IJ}$ encode transport properties in an external magnetic field. Lattice versions of such systems commonly emerge in condensed matter physics in the presence of magnetic fields, including the famed `Hofstadter butterfly' \cite{butterfly}. In our context, the most striking aspect of this generalization is that it may lead to all-bands-flat systems \cite{creutzlddr,allflat}, where 
the spectral singularities of the Laplacian start looking zero-dimensional since there is no dispersion, with all the implications that has for SSB, while the spatial geometry is that of a higher-dimensional lattice. Of course, products of such all-bands-flat lattices and regular lattices, or more elaborate structures, could be considered, with the corresponding effects on
the magnetic Laplacian spectra.

Overall, from concrete examples, it appears that a number of scenarios exist where low spectral dimensions and the corresponding spectral singularities
emerge from natural network constructions. It would be valuable to develop a more comprehensive mathematical theory that controls the strength of these
singularities near the origin in the spectral density of graph Laplacians.


\section{Conclusions}

We have revisited the topic of spontaneous symmetry breaking (SSB) in quantum field theories and its absence in low spatial dimensions, and explored how the issue is illuminated by first discretizing the space and then moving on to a greater variety of discrete geometries.

In application to ordinary field theories, this approach confers the advantage of completely avoiding both ultraviolet and infrared divergences. The ultraviolet divergences are removed by the spatial discretization. The infrared singularities are removed by considering finite-sized discrete spaces, and then examining
how the statistics of the ground state in field space depends on the size of the background geometry. In higher dimensions, the field fluctuations remain finite as the background geometry expands, leaving a preferred field value that signifies that the field shift symmetry is spontaneously broken. In low dimensions, the fluctuations diverge and the symmetry is restored. We thus rederive the classic results on the absence of SSB in low dimensions---as well as their contemporary generalizations to higher-derivative theories---entirely in the language of undergraduate-level quantum mechanics of lattices of oscillators.

Once we allow arbitrary discrete geometries, a much greater variety of backgrounds is possible, and SSB is controlled by a generalization of the resistance distance, and on-average over-the-graph by the usual spectral dimension. A number of constructions for networks with tunable spectral dimensions
have been reported in the literature, creating an array of options for networks that support or prevent SSB. In this regard, a more systematic mathematical understanding of the singularities near zero in the eigenvalue densities of graph Laplacians, which control the spectral dimension, would be highly desirable.


\section*{Acknowledgments}

I thank Ignatios Antoniadis for discussions on spontaneous symmetry breaking in de Sitter,  Jarah Evslin
for multiple discussions on ground states of low-dimensional QFTs and of Gaussian distributions for onsite field values in the vacuum, 
Peter Sollich for answering a few questions about reference \cite{heterogeneous},
Alessandro Vezzani for bringing to my attention the older works dealing with the Mermin-Wagner theorem on graphs,
Sergej Flach for a discussion on the all-bands-flat lattices, and the anonymous referees for helping me with establishing
better connections to the contemporary network science literature.
This work is supported by the  C2F program at Chulalongkorn University and by NSRF via grant number B41G680029,
and additionally by a Priority Research
Area DigiWorld grant under the Strategic Programme
Excellence Initiative at the Jagiellonian University
(Krak\'ow, Poland).



\begin{thebibliography}{99}

\bibitem{Nambu}
Y.~Nambu,
{\it Nobel Lecture: Spontaneous symmetry breaking in particle physics:\\ a case of cross fertilization,}
\doi{Rev.\ Mod.\ Phys. {\bf 81} (2009) 1015}{10.1103/RevModPhys.81.1015}.

\bibitem{Kibble}
T.~W.~B.~Kibble,
{\it Englert-Brout-Higgs-Guralnik-Hagen-Kibble mechanism (history),}\\
\doi{Int. J. Mod. Phys. A \textbf{24} (2009) 6001}{10.1142/S0217751X09048137}.

\bibitem{Englert}
F.~Englert, {\it Nobel Lecture: the BEH mechanism and its scalar boson,}\\
\doi{\it Rev.\ Mod.\ Phys. {\bf 86} (2014) 843}{10.1103/RevModPhys.86.843}.

\bibitem{Weinberg}
S.~Weinberg, {\it Nobel Lecture: Conceptual foundations of the unified theory of weak and\\ electromagnetic interactions,}
\doi{Rev.\ Mod.\ Phys. {\bf 52} 1980) 515 }{10.1103/RevModPhys.52.515}.

\bibitem{MW}
N.~D.~Mermin and H.~Wagner,
{\it Absence of ferromagnetism or antiferromagnetism\\ in one-dimensional or two-dimensional isotropic Heisenberg models,}\\
\doi{Phys. Rev. Lett. \textbf{17} (1966) 1133}{10.1103/PhysRevLett.17.1133}.

\bibitem{Hohenberg}
P.~C.~Hohenberg,
{\it Existence of long-range order in one and two dimensions,}\\
\doi{Phys. Rev. \textbf{158} (1967) 383}{10.1103/PhysRev.158.383}.

\bibitem{Coleman}
S.~R.~Coleman,
{\it There are no Goldstone bosons in two-dimensions,}\\
\doi{Comm. Math. Phys. \textbf{31} (1973) 259}{10.1007/BF01646487}.

\bibitem{MaRajaraman}
S.-k.~Ma and R.~Rajaraman,
{\it Comments on the absence of spontaneous symmetry breaking in low dimensions,}
\doi{Phys. Rev. D \textbf{11} (1975) 1701}{10.1103/PhysRevD.11.1701}.

\bibitem{Ratra}
B.~Ratra,
{\it Restoration of spontaneously broken continuous symmetries in de Sitter\\ spacetime,}
\doi{Phys. Rev. D \textbf{31} (1985) 1931}{10.1103/PhysRevD.31.1931}.

\bibitem{dSinst}
I.~Antoniadis, J.~Iliopoulos and T.~N.~Tomaras,
{\it Quantum instability of de Sitter space,}\\
\doi{Phys. Rev. Lett. \textbf{56} (1986) 1319}{10.1103/PhysRevLett.56.1319}.

\bibitem{dScosmo}
I.~Antoniadis, P.~O.~Mazur and E.~Mottola,
{\it Cosmological dark energy: prospects for a dynamical theory,}
\doi{New J. Phys. \textbf{9} (2007) 11}{10.1088/1367-2630/9/1/011},
\arXiv{gr-qc/0612068}.

\bibitem{Lukyanov}
S.~L.~Lukyanov,
{\it Form-factors of exponential fields in the sine-Gordon model,}\\
\doi{Mod. Phys. Lett. A \textbf{12} (1997), 2543}{10.1142/S0217732397002673},
\arXiv{hep-th/9703190}.

\bibitem{RychkovVitale}
S.~Rychkov and L.~G.~Vitale,
{\it Hamiltonian truncation study of the $\varphi^4$ theory in two dimensions,}
\doi{Phys. Rev. D \textbf{91} (2015) 085011}{10.1103/PhysRevD.91.085011},
\arXiv{1412.3460} [hep-th];
{\it Hamiltonian truncation study of the $\varphi^4$ theory in two dimensions II: the $\mathbb Z_2$ -broken phase and the Chang duality,}
\doi{Phys. Rev. D \textbf{93} (2016) 065014}{10.1103/PhysRevD.93.065014},
\arXiv{1512.00493} [hep-th].

\bibitem{Brauner}
T.~Brauner,
{\it Spontaneous symmetry breaking and Nambu-Goldstone bosons in quantum many-body systems,}
\doi{Symmetry \textbf{2} (2010) 609}{10.3390/sym2020609},
\arXiv{1001.5212} [hep-th].

\bibitem{WB}
H.~Watanabe and T.~Brauner,
{\it On the number of Nambu-Goldstone bosons and its relation to charge densities,}
\doi{Phys. Rev. D \textbf{84} (2011) 125013}{10.1103/PhysRevD.84.125013},
\arXiv{1109.6327} [hep-ph].

\bibitem{Kapustin}
A.~Kapustin,
{\it Remarks on nonrelativistic Goldstone bosons,}
\arXiv{1207.0457} [hep-ph].

\bibitem{WM}
H.~Watanabe and H.~Murayama,
{\it Effective Lagrangian for nonrelativistic systems,}\\
\doi{Phys. Rev. X \textbf{4} (2014)  031057}{10.1103/PhysRevX.4.031057},
\arXiv{1402.7066} [hep-th].

\bibitem{GGHY}
T.~Griffin, K.~T.~Grosvenor, P.~Ho\v{r}ava and Z.~Yan,
{\it Scalar field theories with polynomial shift symmetries,}
\doi{Comm.\ Math.\ Phys. \textbf{340} (2015) 985}{10.1007/s00220-015-2461-2},
\arXiv{1412.1046} [hep-th];\\ 
{\it Cascading multicriticality in nonrelativistic spontaneous symmetry breaking,}\\
\doi{Phys. Rev. Lett. \textbf{115} (2015) 241601}{10.1103/PhysRevLett.115.241601}
\arXiv{1507.06992} [hep-th].

\bibitem{Creutz}M.~Creutz, {\it\doi{Quarks, gluons and lattices}{10.1017/9781009290395}} (CUP, 2023).

\bibitem{Nepomechie}
R.~I.~Nepomechie,
{\it Approaches to a non-abelian antisymmetric tensor gauge field theory,}
\doi{Nucl. Phys. B \textbf{212} (1983) 301}{10.1016/0550-3213(83)90306-1}.

\bibitem{CS}
T.~Jacobson and T.~Sulejmanpasic,
{\it Modified Villain formulation of Abelian Chern-Simons theory,}
\doi{Phys. Rev. D \textbf{107} (2023) 125017}{10.1103/PhysRevD.107.125017},
\arXiv{2303.06160} [hep-th].

\bibitem{randlat}
N.~H.~Christ, R.~Friedberg and T.~D.~Lee,
{\it Random lattice field theory: general formulation,}
\doi{Nucl. Phys. B \textbf{202} (1982) 89}{10.1016/0550-3213(82)90222-X};
{\it Weights of links and plaquettes in a random lattice,}
\doi{Nucl. Phys. B \textbf{210} (1982) 337}{10.1016/0550-3213(82)90124-9}.

\bibitem{Loll}
R.~Loll,
{\it Quantum gravity from causal dynamical triangulations: a review,}
\doi{ Class. Quant. Grav. \textbf{37} (2020) 013002}{10.1088/1361-6382/ab57c7},
\arXiv{1905.08669} [hep-th].

\bibitem{smpl}
R.~C.~Brower, M.~Cheng, G.~T.~Fleming, A.~D.~Gasbarro, T.~G.~Raben, C.~I.~Tan and E.~S.~Weinberg,
{\it Lattice $\phi^4$ field theory on Riemann manifolds: numerical tests for the 2-d Ising CFT on $\mathbb{S}^2$,}
\doi{Phys. Rev. D \textbf{98} (2018) 014502}{10.1103/PhysRevD.98.014502},
\arXiv{1803.08512} [hep-lat].

\bibitem{cdt1}
A.~Candido, G.~Clemente, M.~D'Elia and F.~Rottoli,
{\it Compact gauge fields\\ on causal dynamical triangulations: a 2d case study,}
\doi{JHEP \textbf{04} (2021) 184}{10.1007/JHEP04(2021)184},\\
\arXiv{2010.15714} [hep-lat].

\bibitem{cdt2}
J.~Ambj{\o}rn, Z.~Drogosz, J.~Gizbert-Studnicki, A.~G{\"o}rlich, J.~Jurkiewicz and D.~N{\'e}meth,
{\it Scalar fields in causal dynamical triangulations,}
\doi{Class. Quant. Grav. \textbf{38} (2021) 195030}{10.1088/1361-6382/ac2135},\\
\arXiv{2105.10086} [gr-qc].

\bibitem{Filk}
T.~Filk,
{\it Random graph gauge theories as toy models for nonperturbative string theories,}
\doi{Class. Quant. Grav. \textbf{17} (2000) 4841}{10.1088/0264-9381/17/23/304},
\arXiv{hep-th/0010126}.

\bibitem{qft1}
G.~Bianconi,
{\it The topological Dirac equation of networks and simplicial complexes,}\\
\doi{J. Phys. Compl. {\bf 2} (2021) 035022}{10.1088/2632-072X/ac19be},
\arXiv{2106.02929} [cond-mat.dis-nn].

\bibitem{qft2}
G.~Bianconi,
{\it Dirac gauge theory for topological spinors in 3+1 dimensional networks,}\\
\doi{J. Phys. A \textbf{56} (2023) no.27, 275001}{10.1088/1751-8121/acdc6a},
\arXiv{2212.05621} [cond-mat.dis-nn].

\bibitem{qft3}
N.~Delporte, S.~Sen and R.~Toriumi,
{\it Dirac walks on regular trees,}\\
\doi{J. Phys. A \textbf{57} (2024) 275002}{10.1088/1751-8121/ad4d2e},
\arXiv{2312.10881} [cond-mat.stat-mech].

\bibitem{qft4}
G.~Bianconi,
{\it Quantum entropy couples matter with geometry,}\\
\doi{J. Phys. A \textbf{57} (2024) 365002}{10.1088/1751-8121/ad6f7e},
\arXiv{2404.08556} [cond-mat.dis-nn].

\bibitem{qft5}
R.~Wang, Y.~Tian, P.~Li{\`o} and G.~Bianconi,
{\it Dirac-equation signal processing:\\ physics boosts topological machine learning,}
\doi{PNAS Nexus \textbf{4} (2025) pgaf139}{10.1093/pnasnexus/pgaf139},\\
\arXiv{2412.05132} [cond-mat.dis-nn].

\bibitem{netmass}G.~Bianconi, {\it The mass of simple and higher-order networks,}
\doi{J.\ Phys. A {\bf 57} (2024) 015001}{10.1088/1751-8121/ad0fb5},
\arXiv{2309.07851} [cond-mat.dis-nn].

\bibitem{combi}
C.~A.~Trugenberger,
{\it Combinatorial quantum gravity: geometry from random bits,}\\
\doi{JHEP \textbf{09} (2017) 045}{10.1007/JHEP09(2017)045},
\arXiv{1610.05934} [hep-th].

\bibitem{birth}
P.~Akara-pipattana, T.~Chotibut and O.~Evnin,
{\it The birth of geometry in exponential\\ random graphs,}
\doi{J. Phys. A \textbf{54} (2021) 425001}{10.1088/1751-8121/ac2474},
\arXiv{2102.11477} [cond-mat.dis-nn].

\bibitem{combirev}
C.~A.~Trugenberger,
{\it Networks as the fundamental constituents of the universe,}\\
 \doi{J.\ Phys.\ Complex. {\bf 6} (2025) 042001}{10.1088/2632-072X/ae29d3},
\arXiv{2512.17676} [gr-qc].

\bibitem{Hamiltonian}
J.~B.~Kogut and L.~Susskind,
{\it Hamiltonian formulation of Wilson's lattice gauge theories,}
\doi{Phys. Rev. D \textbf{11} (1975) 395}{10.1103/PhysRevD.11.395}.
\arXiv{hep-th/0505113}.

\bibitem{Kittel}Ch.~Kittel, {\it Introduction to solid state physics} (John Wiley \& Sons, 2005).

\bibitem{latticecount}L.~Parnovski and N.~Sidorova.
{\it Critical dimensions for counting lattice points in\\ Euclidean annuli,}
\doi{Math.\ Mod.\ Nat.\ Phen. {\bf 5} (2010) 293}{10.1051/mmnp/20105413}.

\bibitem{HOVHS}
L.~Classen and J.~J.~Betouras,
{\it High-order Van Hove singularities and their\\ connection to flat bands,}
\doi{Ann.\ Rev.\ Cond.\ Mat.\ Phys. {\bf 16} (2025) 229}{10.1146/annurev-conmatphys-042924-015000},\\
\arXiv{2405.20226} [cond-mat.str-el].

\bibitem{Rajaraman}
R.~Rajaraman, {\it Solitons and instantons} (North Holland, 1982).

\bibitem{thesis}
O.~Evnin,
{\it On quantum interacting embedded geometrical objects of various dimensions,}\\
\href{https://inspirehep.net/files/6ee200f06b1f630c9fb6d6acfa6045f0}{Caltech PhD thesis} (2006).

\bibitem{localr}
B.~Craps, O.~Evnin and S.~Nakamura,
{\it Local recoil of extended solitons:\\ a string theory example,}
\doi{JHEP \textbf{01} (2007) 050}{10.1088/1126-6708/2007/01/050},
\arXiv{hep-th/0608123}.

\bibitem{EvslinLiu}
J.~Evslin and H.~Liu,
{\it An extended soliton's zero modes,}\\
\doi{JHEP \textbf{12} (2025) 015}{10.1007/JHEP12(2025)015},
\arXiv{2507.18922} [hep-th].

\bibitem{mesonsoli}
M.~Uehara, A.~Hayashi and S.~Saito,
{\it Meson-soliton scattering with full recoil in standard collective coordinate quantization,}
\doi{Nucl. Phys. A \textbf{534} (1991) 680}{10.1016/0375-9474(91)90466-J}.

\bibitem{D0r}
B.~Craps, O.~Evnin and S.~Nakamura,
{\it D0-brane recoil revisited,}\\
\doi{JHEP \textbf{12} (2006) 081}{10.1088/1126-6708/2006/12/081},
\arXiv{hep-th/0609216}.

\bibitem{EvslinGuo}
J.~Evslin and H.~Guo,
{\it Alternative to collective coordinates,}\\
\doi{Phys. Rev. D \textbf{103} (2021) L041701}{10.1103/PhysRevD.103.L041701},
\arXiv{2101.08028} [hep-th].

\bibitem{SenStefanski}
A.~Sen and B.~Stefa\'nski Jr.,
{\it Scattering of D0-branes and strings,}\\
\doi{JHEP \textbf{01} (2026) 033}{10.1007/JHEP01(2026)033},
\arXiv{2509.02716} [hep-th].

\bibitem{plates}A.~W.~Leissa, {\it Vibration of plates} (NASA, 1969).

\bibitem{laprenorm}P.~Villegas, T.~Gili, G.~Caldarelli and A.~Gabrielli, {\it Laplacian\\ renormalization group for heterogeneous networks,}
\doi{Nat.\ Phys. {\bf 19} (2023) 445}{10.1038/s41567-022-01866-8},\\ \arXiv{2203.07230} [cond-mat.stat-mech].

\bibitem{KleinRandic}
D.~J.~Klein and  M.~Randi\'c, {\it Resistance distance,} \doi{J.\ Math.\ Chem. {\bf 12}  (1993) 81}{10.1007/BF01164627}.

\bibitem{resist}
P.~Akara-pipattana, T.~Chotibut and O.~Evnin,
{\it Resistance distance distribution\\ in large sparse random graphs,}
\doi{J. Stat. Mech. \textbf{2203} (2022) 033404}{10.1088/1742-5468/ac57ba},\\
\arXiv{2107.12561} [cond-mat.dis-nn].

\bibitem{LPS}C.~R.~Laumann, S.~A.~Parameswaran and S.~L.~Sondhi,
{\it Absence of Goldstone bosons on the Bethe lattice,}
\doi{Phys.\ Rev. B {\bf 80} (2009) 144415}{10.1103/PhysRevB.80.144415},
\arXiv{0906.5098} [cond-mat.stat-mech].

\bibitem{Dhar}
D.~Dhar, {\it Lattices of effectively nonintegral dimensionality,}\\
\doi{J.\ Math.\ Phys. {\bf 18} (1977) 577}{10.1063/1.523316}.

\bibitem{HughesShlesinger}
B.~D.~Hughes and M.~F.~Shlesinger,
{\it Lattice dynamics, random walks, and nonintegral\\ effective dimensionality,}
\doi{J.\ Math.\ Phys. {\bf 23} (1982) 1688}{10.1063/1.525554}.

\bibitem{AlexanderOrbach}
S.~Alexander and R.~Orbach, {\it Density of states on fractals:  ``fractons'',}\\
\doi{J.\ Physique Lett. {\bf 43} (1982) 625}{10.1051/jphyslet:019820043017062500}

\bibitem{CassiRegina}
D.~Cassi and S.~Regina,
{\it Spectral dimension of branched structures: universality\\ in geometrical disorder,}
\doi{Phys.\ Rev.\ Lett. {\bf 70} (1993)1647}{10.1103/PhysRevLett.70.1647}.

\bibitem{BurioniCassi}
R.~Burioni and D.~Cassi,
{\it Universal properties of spectral dimension,}\\
\doi{Phys.\ Rev.\ Lett. {\bf 76} (1996) 1091}{10.1103/PhysRevLett.76.1091}.

\bibitem{universe}
J.~Ambj\o rn, J.~Jurkiewicz and R.~Loll,
{\it The spectral dimension of the universe is scale dependent,}
\doi{Phys. Rev. Lett. \textbf{95} (2005) 171301}{10.1103/PhysRevLett.95.171301},
\arXiv{hep-th/0505113}.

\bibitem{specdim1}A.~P.~Mill\'an, J.~J.~Torres and G.~Bianconi, {\it Synchronization in network\\ geometries with finite spectral dimension,}
\doi{Phys.\ Rev. E {\bf 99} (2019) 022307}{10.1103/PhysRevE.99.022307},\\
\arXiv{1811.03069} [cond-mat.dis-nn].

\bibitem{specdim2}J.~J.~Torres and G.~Bianconi, {\it Simplicial complexes: higher-order spectral dimension and dynamics,}
\doi{J.\ Phys.\ Complex. {\bf 1} (2020) 015002}{10.1088/2632-072X/ab82f5},
\arXiv{2001.05934} [cond-mat.dis-nn].

\bibitem{specdim3}G.~Bianconi and S.~N.~Dorogovtsev, {\it The spectral dimension of simplicial complexes: a renormalization group theory,}
\doi{J.\ Stat.\ Mech. {\bf 2020} (2020) 014005}{10.1088/1742-5468/ab5d0e},
\arXiv{1910.12566} [cond-mat.dis-nn].

\bibitem{ssb1}D.~Cassi, {\it Phase transitions and random walks on graphs: A generalization of the Mermin-Wagner theorem to disordered lattices, fractals, and other discrete structures,} \doi{Phys.\ Rev.\ Lett. {\bf 68} (1992) 3631}{10.1103/PhysRevLett.68.3631}; {\it Local vs average behavior on inhomogeneous structures: recurrence on the average and a further extension of Mermin-Wagner theorem on graphs,} \doi{Phys.\ Rev.\ Lett. {\bf 76} (1996) 2941}{10.1103/PhysRevLett.76.2941}.

\bibitem{ssb2}R.~Burioni, D.~Cassi and A.~Vezzani, {\it Inverse Mermin-Wagner theorem for classical spin models on graphs,}
\doi{Phys.\ Rev. E {\bf 60} (1999) 1500}{10.1103/PhysRevE.60.1500}.

\bibitem{sph}
D.~Cassi and L.~Fabbian, {\it The spherical model on graphs,}
\doi{J.\ Phys. A {\bf 32} (1999) L93}{10.1088/0305-4470/32/8/001}.

\bibitem{infsph}R.~Burioni, D.~Cassi and C.~Destri, {\it $n\to\infty$ limit of $O(n)$ ferromagnetic models on graphs,}
 \doi{Phys.\ Rev.\ Lett. {\bf 85} (2000) 1496}{10.1103/PhysRevLett.85.1496}.

\bibitem{RB}A.~J.~Bray and G.~J.~Rodgers, {\it Diffusion in a sparsely connected space:\\ a model for glassy relaxation},
\doi{Phys.\ Rev.\ B {\bf 38} (1988) 11461}{10.1103/PhysRevB.38.11461}.

\bibitem{onthem}
A.~N.~Samukhin, S.~N.~Dorogovtsev and J.~F.~F.~Mendes, {\it Laplacian spectra of complex networks and random walks on them: Are scale-free architectures really important?,}\\ \doi{Phys.\ Rev.\ E {\bf 77} (2008) 036115}{10.1103/PhysRevE.77.036115}, \arXiv{0706.1176} [cond-mat.stat-mech].

\bibitem{laplace}P.~Akara-pipattana and O.~Evnin,
{\it Random matrices with row constraints and eigenvalue distributions of graph Laplacians,}
\doi{J. Phys. A \textbf{56} (2023) 295001}{10.1088/1751-8121/acdcd3},
\arXiv{2212.06499} [cond-mat.dis-nn]; {\it Hammerstein equations for sparse random matrices,}
\doi{J.\ Phys. A \textbf{58} (2025) 035006}{10.1088/1751-8121/ada8ea},
\arXiv{2410.00355} [cond-mat.dis-nn].

\bibitem{Sierpinski}
L.~De Carli, A.~Echezabal and I.~Morell, {\it On the Sierpi\'nski triangle and its generalizations,}
\arXiv{2506.20456} [math.NT].

\bibitem{Strichartz}
R.~S.~Strichartz,
{\it Analysis on fractals,}
\href{https://www.ams.org/notices/199910/fea-strichartz.pdf}{Not.\ Am.\ Math.\ Soc. {\bf 46} (1999) 1199}.

\bibitem{StrichartzTeplyaev}
R.~Strichartz and A.~Teplyaev, {\it Spectral analysis on infinite Sierpi\'nski fractafolds,}\\
\doi{J.\ Anal.\ Math. {\bf 116} (2012) 255}{10.1007/s11854-012-0007-5},
\arXiv{1011.1049} [math.FA].

\bibitem{HilferBlumen}
R.~Hilfer and A.~Blumen,
{\it Renormalisation on Sierpinski-type fractals,}\\
\doi{J.\ Phys. A {\bf 17} (1984) L537}{10.1088/0305-4470/17/10/004}.

\bibitem{Dunne}
G.~V.~Dunne,
{\it Heat kernels and zeta functions on fractals,}
\doi{J. Phys. A \textbf{45} (2012) 374016}{10.1088/1751-8113/45/37/374016},\\
\arXiv{1205.2723} [math-ph].

\bibitem{Hill}
C.~T.~Hill,
{\it Fractal theory space: space-time of noninteger dimensionality,}\\
\doi{Phys. Rev. D \textbf{67} (2003) 085004}{10.1103/PhysRevD.67.085004},
\arXiv{hep-th/0210076}.

\bibitem{nested}
T.~Ni and Z.~Wen,
{\it Construction of free quantum fields on nested fractal space-times,}\\
\doi{J. Math. Phys. \textbf{64} (2023) 122303}{10.1063/5.0147003}.

\bibitem{ACAT1}R.~Abou-Chacra, D.~J.~Thouless  and P.~W.~Anderson, {\it A selfconsistent theory of localization,} \doi{J.\ Phys. C {\bf 6} (1973) 1734}{10.1088/0022-3719/6/10/009}.

\bibitem{ACAT2}R.~Abou-Chacra and D.~J.~Thouless, {\it Self-consistent theory of localization II: localization near the band edges,} \doi{J.\ Phys. C {\bf 7} (1974) 65}{10.1088/0022-3719/7/1/015}. 

\bibitem{nuclphysb}A.~D.~Mirlin and Y.~V.~Fyodorov, {\it Localization transition in the Anderson model\\ on the Bethe lattice: spontaneous symmetry breaking and correlation functions,}\\ \doi{Nucl.\ Phys. B {\bf366} (1991) 507}{10.1016/0550-3213(91)90028-V}.

\bibitem{tunable}
A.~P.~Mill\'an, G.~Gori, F.~Battiston, T.~Enss and  N.~Defenu, {\it Complex networks with tuneable dimensions as a universality playground,}
\doi{Phys.\ Rev. Res. {\bf 3} (2021) 023015}{10.1103/PhysRevResearch.3.023015},\\
\arXiv{2006.10421} [cond-mat.stat-mech].

\bibitem{smallw}
D.~J.~Watts and S.~H.~Strogatz,  {\it Collective dynamics of `small-world' networks,}\\
 \doi{Nature {\bf 393} (1998) 440}{10.1038/30918}.

\bibitem{heterogeneous}
J.~D.~da Silva, D.~Tapias, P.~Sollich and F.~L.~Metz,
{\it Spectral properties, localization\\ transition and multifractal eigenvectors of the Laplacian on heterogeneous networks,}\\
\doi{SciPost Phys. \textbf{18} (2025) 047}{10.21468/SciPostPhys.18.2.047}, \arXiv{2408.13322} [cond-mat.dis-nn].

\bibitem{molinari}
L.~G.~Molinari, {\it Non-Hermitian spectra and Anderson localization,}\\
\doi{J. Phys. A {\bf 42} (2009) 265204}{10.1088/1751-8113/42/26/265204}, \arXiv{0808.1241} [math-ph].

\bibitem{magnetic}
Ch.~Hu, B.~Hua, S.~Kamtue, Sh.~Liu, F.~Münch and N.~Peyerimhoff,\\
{\it On a magneto-spectral invariant on finite graphs,}
\arXiv{2507.03754} [math.SP].

\bibitem{butterfly}J.~G.~Analytis, S.~J.~Blundell and A.~Ardavan,
{\it Landau levels, molecular orbitals,\\ and the Hofstadter butterfly in finite systems,}
\doi{Am.\ J.\ Phys. 72 (2004) 613}{10.1119/1.1615568}.

\bibitem{creutzlddr}
M.~Creutz,
{\it End states, ladder compounds, and domain wall fermions,}\\
\doi{Phys. Rev. Lett. \textbf{83} (1999) 2636}{10.1103/PhysRevLett.83.2636},
\arXiv{hep-lat/9902028}.

\bibitem{allflat}T.~\v Cade\v z, Y.~Kim, A.~Andreanov and S.~Flach, {\it Metal-insulator transition\\ in infinitesimally weakly disordered flatbands,}
\doi{Phys.\ Rev. B {\bf 104} (2021) L180201}{10.1103/PhysRevB.104.L180201},\\
\arXiv{2107.11365} [cond-mat.dis-nn].

\end{thebibliography}
\end{document}